# Interference Evidence for Rashba-Type Spin-Split on Semimetallic WTe$_2$ Surface

Qing Li[1], Jiaqiang Yan[2,3], Biao Yang[1], Yunyi Zang[4], Junjie Zhang[1], Ke He[4], Menghao Wu[5], Yanfei Zhao[6], David Mandrus[3,4], Jian Wang[6], Qikun Xue[4], Lifeng Chi[1]*, David J. Singh[7],* and Minghu Pan[5],*

[1]Institute of Functional Nano and Soft Materials (FUNSOM) and Collaborative Innovation Center of Suzhou Science and Technology, Soochow University, Jiangsu 215123, China.

[2]Department of Materials Science and Engineering, University of Tennessee, Knoxville, Tennessee 37996, USA.

[3]Materials Science and Technology Division, Oak Ridge National Laboratory, Oak Ridge, Tennessee 37831, USA.

[4]State Key Laboratory of Low-Dimensional Quantum Physics, Department of Physics, Tsinghua University, Beijing 100084, China

[5]School of Physics, Huazhong University of Science and Technology, Wuhan 430074, China.

[6]International Center for Quantum Materials, School of Physics, Peking University, Beijing 100871, China

[7]Department of Physics and Astronomy, University of Missouri, Columbia, Missouri 65211-7010, USA

*e-mail: mhupan@gmail.com; chilf@suda.edu.cn; singhdj@missouri.edu

Semimetallic tungsten ditelluride ($WTe_2$) displays an extremely large non-saturating magnetoresistance (XMR), which is the subject of intense interest. This phenomenon is thought to arise from the combination of perfect n-p charge compensation with low carrier densities in $WTe_2$ and presumably details of its band structure. Recently, "spin texture" induced by strong spin-orbital coupling (SOC) has been observed in $WTe_2$ by angle-resolved photoemission spectroscopy (ARPES). This provides a mechanism for protecting backscattering for the states involved and thus was proposed to play an important role in the XMR of $WTe_2$. Here, based on our density functional calculations for bulk $WTe_2$, we found a strong Rashba spin-orbit effect in the calculated band structure due to its non-centrosymmetric structure. This splits bands and two-fold spin degeneracy of bands is lifted. A prominent Umklapp interference pattern (a spectroscopic feature with involving reciprocal lattice vectors) can be observed by scanning tunneling microscopic (STM) measurements on $WTe_2$ surface at 4.2 K. This differs distinctly from the surface atomic structure demonstrated at 77 K. The energy dependence of Umklapp interference shows a strong correspondence with densities of states integrated from ARPES measurement, manifesting a fact that the bands are spin-split on the opposites side of Γ. Spectroscopic survey reveals the ratio of electron/hole asymmetry changes alternately with lateral locations along b axis, providing a microscopic picture for double-carrier transport of semimetallic $WTe_2$. The calculated band structure and Fermi surface is further supported by our

ARPES results and Shubnikov-de Haas (SdH) oscillations measurements.

WTe$_2$, a layered transition-metal dichalcogenide (TMD), whose sheets consist of a tungsten layer sandwiched by adjacent chalcogenide layers, has been the subject of intense interest for its extraordinary magnetoresistance (XMR)[1-6]. The non-saturation of the resistance with magnetic field differs significantly with the normal MR phenomenon, where the magnetoresistance is quadratic only in low fields, and tends to saturate at high fields[1]. This phenomenon has been ascribed to the perfect n-p charge compensation in WTe$_2$, based on the investigations by angle-resolved photoelectron spectroscopy (ARPES)[7] and theoretical calculations[3]. According to the two-band model for semimetals, the sharp resonance with exact compensation (n=p) results in that MR given by $\Delta\rho/\rho=\mu^2 B^2$ and $\rho_{xx}$ grows with $B^2$ without saturation[1,3]. Here, µ is the average mobility of the carriers.

On the other hand, tungsten chains are formed within dichalcogenide layers along the *a* axis of the WTe$_2$ orthorhombic unit cell, making the compound structurally one dimensional. As a consequence, a quasi-one dimensional, semi-metallic Fermi surface topology of WTe$_2$ along the Γ-X direction (corresponding to the *a*-axis, along the tungsten chains in real space) was found[1]. Very recently, a complicated Fermi surface of WTe$_2$ was reported via the ARPES measurements[8], in which nine Fermi pockets were resolved, including one hole pocket around the Brillouin zone center, and two hole pockets and two electron pockets on each side of Γ point along the Γ-X direction. This observation differs from the simple Fermi surface topology (one electron and one hole pocket) significantly. More important, similar to the case of topological insulators[9-11], the circular dichroism, which is a signature for strong spin-orbital coupling (SOC) and a spin texture, was observed at the Fermi surface in photoemission spectra[8]. Such spin texture provides a mechanism to protect the backscattering at zero field thus may be responsible for the XMR.

Here we study in detail the non-centrosymmetric structure and the induced Rashba spin-orbit interactions in $WTe_2$ by density-functional theory (DFT) calculations using a crystal structure in which the atomic positions are determined by total energy minimization. The Rashba effect lifts two-fold spin degeneracy of bands across Fermi level ($E_F$) and produces five Fermi pockets (instead of two normally), including two electron pockets, two hole pockets and one deeper hole band that just kisses $E_F$. In conjunction with scanning tunneling microscopy and spectroscopy (STM/STS) measurements, we observe umklapp-type interference pattern formed on the surface of *in situ* cleaved $WTe_2$ single crystal at 4.2 K. Such an umklapp-type interference pattern is a spectroscopic feature instead of a pure geometric one. It is dominated by spin-conserving processes, with that a reciprocal lattice vector $\vec{g}$ is permitted, whereas ordinary interferences are forbidden by spin conservation at low temperature. The band splitting is further evidenced by the energy dependence of the interference patterns. We note that $WTe_2$ has space group #31, Pmn21, which is non-centrosymmetric due to lack of a mirror or glide perpendicular to the *c*-axis and that the reported XMR effect is for field along *c* and current along *a* and goes to zero when the field is perpendicular to *c*.

**Results**

**DFT investigation of the electronic structures**. DFT calculations, with the full-potential linearized augmented plane wave (LAPW) WIEN2k package and the PBE exchange-correlation functional[12], were performed to investigate the WTe$_2$ bulk structure (see methods for details). Atom positions after relaxation are shown in Supplementary Tab. S1 and the relaxed structure is shown in Fig. 1a. There is an interesting connection between WTe$_2$ and MoS$_2$ in that bulk MoS$_2$ is centrosymmetric, while the single layer version is not, which leads to interesting spin-orbit induced valley effects in MoS$_2$[13].

The band dispersions are shown in Fig. 1b and in agreement with previous reports[1,14] by showing WTe$_2$ to be a semimetal with valence and conduction bands that barely cross the Fermi energy at different places in the Brillouin zone along the Γ–X direction. Our calculated band structure is similar on a large scale to that of Ali and co-workers[1], who also included SOC, but differs in significant details presumably due to the structure used. With SOC, the non-centrosymmetric structure leads to the structure inversion-asymmetric (SIA) spin splitting term in Hamiltonian, *i.e.* a Rashba effect[15-17]. The resulting band structure complex because there are multiple bands including dispersive and less dispersive ones involved, but can is seen for example in the shift of the band extrema away from the zone center, and the dumbbell shapes of the hole Fermi surfaces (Fig. 1d). The calculated band structure has five non-degenerate bands crossing the Fermi level, that are two electron bands, two hole bands and one deeper hole band that just kisses $E_F$ to give very tiny hole sections. A closer inspection of the calculated low-lying electronic structure in the Γ–X direction is illustrated in Fig. 1c (right panel), showing that an original set of electron and hole bands with spin degeneracy splits into two sets of electron and hole bands. The Fermi surfaces are shown in Fig. 1d. The two large hole sheets are connected across the zone center, while the two electron sheets have small necks along the $k_z$ direction. Importantly, the low energy structure is not reasonably characterized as one or even two dimensional. The calculated plasma energies in the three crystallographic directions are $\Omega_{p,x}$=1.07 eV, $\Omega_{p,y}$=0.83 eV and $\Omega_{p,z}$=0.62 eV, which would predict a constant scattering time conductivity anisotropy of $\sigma_x/\sigma_z \sim (\Omega_{p,x}/\Omega_{p,z})^2$ of approximately 3 and similarly $\sigma_x/\sigma_z$ of approximately 1.7, *i.e.* a moderately anisotropic three dimensional semimetal.

The five pockets of the Fermi surface (Fig. 1d) lie along the Γ-X direction, which is supported by SdH oscillations measured with perpendicular field (H//c axis). As shown in Fig. S1, there are four major peaks at 89.5 T, 119.4 T, 141.8 T and 156.7 T, which stand for four extremal areas. With the topology found in calculations this means four pockets, since the necks of the electron sheets along $k_z$ are too small to be one of these. In additional, higher harmonic peaks are also observed. According to the Onsager relation $F = (\phi_0/2\pi^2)A_F$, the cross sectional areas of Fermi surface normal to the field are $A_F$=8.54×10$^{-3}$ Å$^{-2}$, 1.14×10$^{-2}$ Å$^{-2}$, 1.35×10$^{-2}$ Å$^{-2}$ and 1.49×10$^{-2}$ Å$^{-2}$, respectively. Assuming a circular cross-section, four Fermi momentums $k_F$ ≈0.052 Å$^{-1}$, 0.06 Å$^{-1}$, 0.066 Å$^{-1}$ and 0.069 Å$^{-1}$ can be estimated, which is in excellent agreement with recent ARPES result[8] and reasonably agrees our DFT calculations. The anisotropic but three dimensional Fermi surfaces shown here underlie the transport and therefore the anisotropic MR property, which has been demonstrated by our recent transport measurements[18].

The ARPES results (left panel of Fig. 1c) resolve three bands all together across $E_F$ along Γ-X, referred as α, β and δ bands (see methods for experimental details, large scale electronic structure of WTe$_2$ along Γ-X is shown in Fig. S2), respectively. The β and δ bands contribute two Fermi crossings along Γ-X and forms the two hole pockets. The α band gives two Fermi crossings and forms an electron pocket. There are two bands located around 80 meV below $E_F$ (the γ band) and around 190 meV below $E_F$ (the ε band). Except for the tiny hole pocket inside of hole dumbells, all four Fermi pockets predicted by our DFT calculations are clearly resolved. This supports the calculated electronic structure and Fermi surface topology, which is different from that originally reported by Ali *et al.*[1].

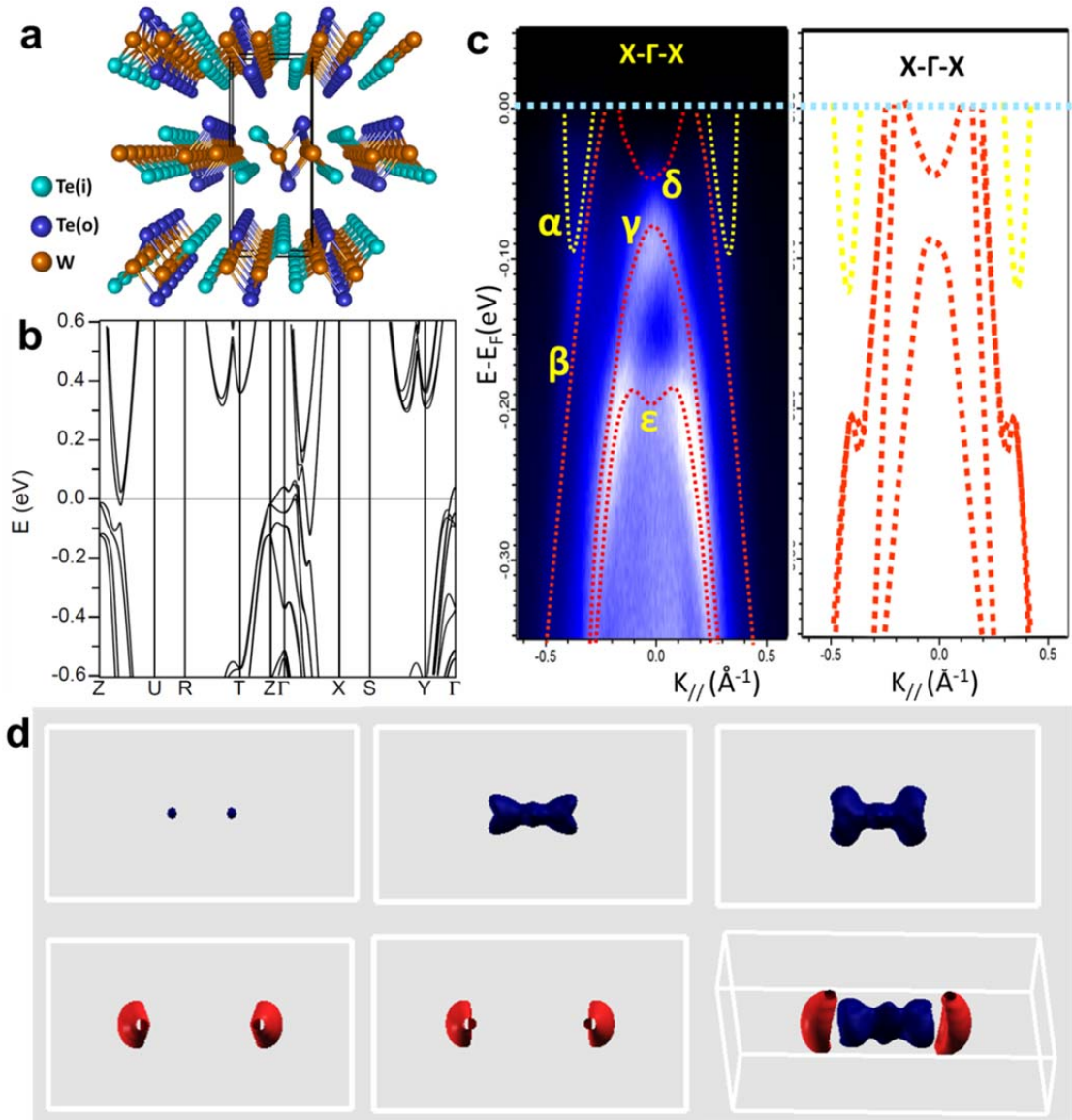

**Figure 1 | DFT calculation of electronic structures of $WTe_2$.** (a) The relaxed bulk structure of $WTe_2$. (b) The calculated band structure of $WTe_2$ bulk, with relaxed structure of the slab. (c) The photoemission intensity plot (left panel) along Γ-X direction together with a closer inspection of the calculated electronic structure (right panel), showing the crossing of the states in the Γ–X direction. The colorful lines are the schematic drawings of the low-lying electronic structure of $WTe_2$, where α, β and δ bands cross at Fermi level while γ and ε are deeper bands below

$E_F$. ARPES data were taken at 100 K with the photon energy 21.2 eV. **(d)** Calculated 3D Fermi surface for WTe$_2$ bulk. Fermi surfaces for the five bands crossing the Fermi level shown as blue for hole surfaces and red for electron surfaces. The bottom right shows the total Fermi surface, but note that only the largest hole and electron surface are visible.

In the topological insulators, SOC induces the spin texture in the topologically protected surface states, namely opposite spin direction on the opposite side of Γ. This provides the protection against backscattering in the absence of magnetic scattering at zero field[19-21]. Such a mechanism may play an important role in the large non-saturating MR of WTe$_2$. Although the existence of spin texture in WTe$_2$ has been detected by ARPES measurements[8], a direct observation of spin-texture and its dependence with external magnetic field are still missing. For this purpose, extensive STM/S measurements were performed (see "Methods" for experimental details).

**STM study of the WTe$_2$ surface.** Surface termination always breaks inversion symmetry. Thus a stronger Rashba effect is expected and affects the Fermi surfaces, screening, and electron dynamics on the surface[22]. Figure 2a shows a representative topographic image of a fresh-cleaved WTe$_2$ surface obtained at 77 K. Based on crystal-chemical considerations and experience with other layered TMDs such as MoSe$_2$ and IrTe$_2$[23,24], cleavage always occurs between adjacent chalcogenide layers so that a Te terminated surface is exposed. The atomic-resolved STM image in Fig. 2a can therefore be understood based on the atomic structure of top Te layer (Top view of the crystal structure of the $WTe_2$ bulk is demonstrated in the left panel of Fig. 2b and the corresponding first Brillouin zone is shown in the right panel of Fig. 2b). On the surface layer of $WTe_2$, the tellurium octahedra distort slightly and the metal atoms are displaced from their ideal octahedral sites, forming zigzag metal-metal chains along *a* axis (Fig. 2b). The Te atom layers are buckled (see Fig. 1a), with the Te(i) atoms displaced a little into the sandwich layer and Te(o) atoms moved slightly out. We therefore assign the brighter feature in Fig. 2a to Te(o) atoms while the darker ones to Te(i) atoms.

Conductance spectra (*dI/dV* versus $V_b$), which measure the local density of states (DOS) at the tip site, were obtained in order to investigate the electronic structure of the $WTe_2$ surface. Thousands tunneling spectra were acquired on clean $WTe_2$ surface at different locations (cyan curves in Fig. 2d) and an averaged *dI/dV* spectrum is also superposed (shown as black curve in Fig. 2d). Figure 2c gives the calculated total DOS for $WTe_2$ bulk and the projections of the electronic density of states (PDOS) onto W (d bands) and Te (p bands) atoms. Both W (d orbitals) and Te (p orbitals) contribute to the overall DOS of valence and conduction bands. The obtained tunneling spectra are in excellent agreement with the theoretical calculations. Without the consideration of the spin, the electronic structure of $WTe_2$ around $E_F$ is fairly simple: The valence band and conduction bands barely cross the Fermi energy at different places in the

Brillouin zone (a schematic scratch of the band structure of WTe$_2$ is shown in the insert of Fig. 2d). We then are able to identify an $E_{CBM}$ (conduction-band minimum) at ∼-40 mV and an $E_{VBM}$ (valence-band maximum) at +40∼+60 mV. The overlap between the valence band and conduction bands is 80 meV to 100 meV, leading to the semimetallic nature of the sample. The spectroscopic results agree reasonably well with our band structure calculations and ARPES measurements[7].

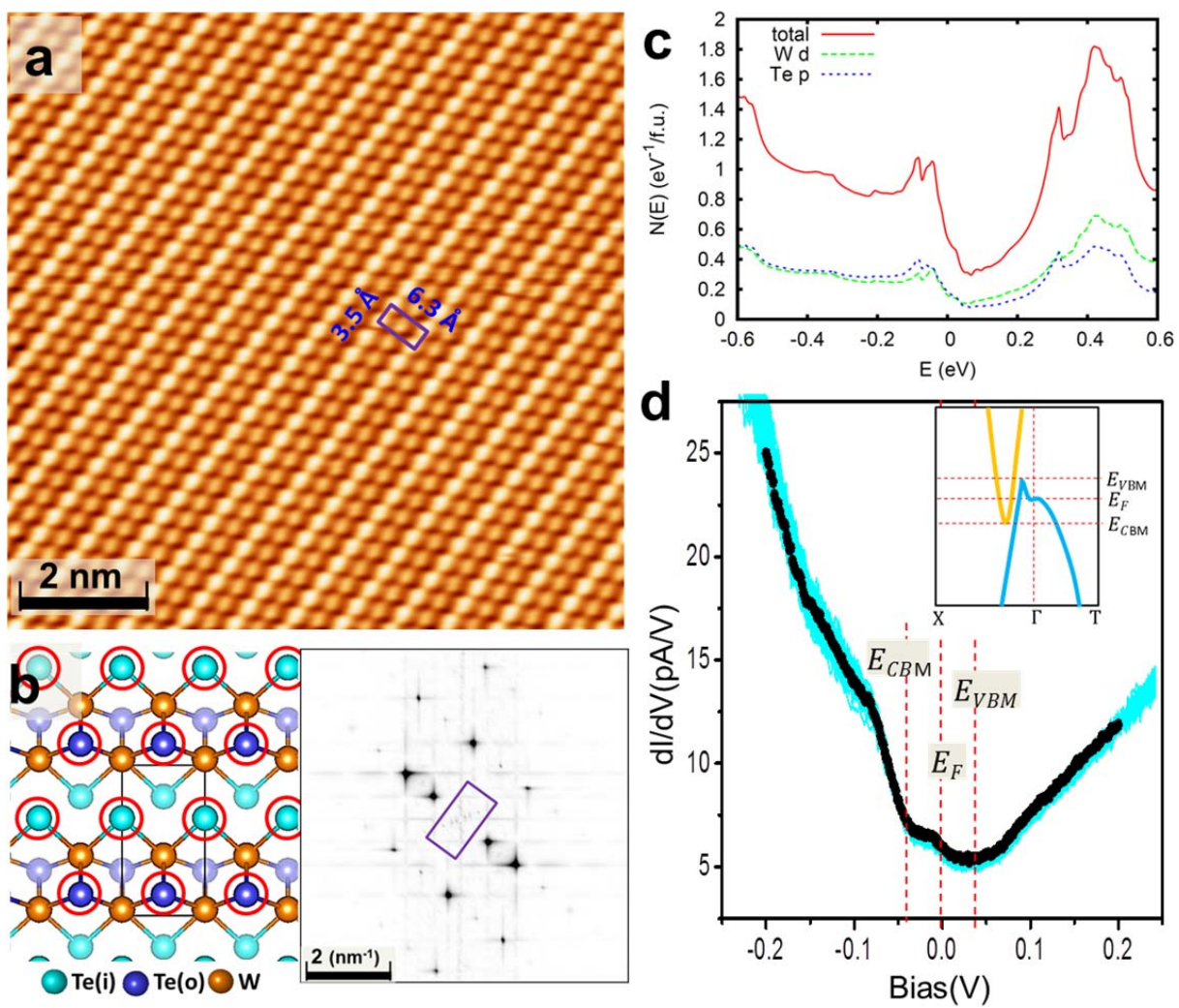

**Figure 2 | STM/S measurements of cleaved WTe$_2$. (a)** Atomically resolved STM topographic image shows the cleaved WTe$_2$ surface. The image was acquired at 77 K. **(b)** Top view of crystal structure of WTe$_2$, with layers of W (golden) atoms sandwiched between layers of Te (blue and cyan) atoms. Here, red circles are used for highlighting top Te(i) and Te(o) atoms, which can be assigned to topographic features observed in atomic resolution STM image in **(a)**. (Right panel) FFT image of a large (~45 nm) atomic resolution STM topographic image, displaying Bragg vectors (dashed red circles) and the corresponding first Brillouin zone (purple rectangle). **(c)** The calculated total DOS, with projections of the DOS onto different atoms, W (d obitals) and Te (p obitals). The energy zero is set to the Fermi level. **(d)** $dI/dV$ spectrum acquired from the surface area in WTe$_2$ at 4.2 K, showing $E_{BCB}$ at ~-40 mV, $E_{TVB}$ at ~+40 mV and $E_F$ at zero. Setpoint parameters: $V_b$ = +50 mV (a), $R_J$ = 0.5 GΩ; $V_{rms}$ = 5 mV. (Inset) Schematic band structure of WTe$_2$ based on our *ab initio* calculations.

Dramatic changes take place when the temperature is lowered to 4.2 K, at which STM images of the surface exhibit a unique "lattice" pattern (Fig. 3a). The unit cell size for this "lattice" pattern is 3.5×6.3 Å, exactly the same as the unit cell of the WTe$_2$ surface and the experimental structure taken at 77 K. The difference between STM topographic images obtained at 77 K and 4.2 K is very obvious. As shown in the left panel of Fig. 3b, the darker protrusions shift from the center (77 K) to the edge (4.2 K) of the unit cell and form a rectangular lattice instead of the triangular one at 77 K. No structural transition has been reported in this temperature range for WTe$_2$. On the other hand, such a structural change for the top Te layer is extremely energetically unfavorable. The rectangular lattice at 4.2 K cannot be attributed to the surface structural/geometric feature, and therefore should represent electronic variations on the surface. Electronic modulations are not rare and have been reported in other materials, mostly arise from a charge ordering. However, charge-density waves (CDW)[25-27] are generally accompanied by at least partial energy gaps in the single-particle excitation spectrum near the Fermi level[28]. Our observed "lattice" pattern cannot be assigned to the charge ordering since no gap opening is obtained in our spectroscopy measurements (Fig. 2d).

Alternatively, this electronic "lattice" pattern can be a result of the quasi-particle interference (QI) induced by surface Rashba SOC splitting. It is vital to realize that states with $\vec{k}_F$ and $-\vec{k}_F$ have opposite spin directions. All of the interference processes building up the FS topology should be forbidden in the absence of spin-flip[29]. One the other hand, in the umklapp-type interference processes, a reciprocal lattice vector $\vec{g}$ are not forbidden by spin. For example, interference between $(\vec{k}, \leftarrow)$ and $(\vec{k}+\vec{g}, \leftarrow)$ is always permitted, but interference between $(\vec{k}, \leftarrow)$ and $(-\vec{k}, \rightarrow)$ is not (Fig. 3b, right panel). As a consequence, this does not require the existence of defect scattering and the pattern is formed extending the entire sample surface where a regular crystalline lattice of $WTe_2$ preserved, as that reported for the Be(110) surface[30].

The umklapp-type interference processes can be further evidenced by differential conductance mapping at different energy levels. The differential conductance maps taken with sample bias varying from -0.4 V to 0.5 V are shown in Supplementary Fig. S3 and a representative *dI/dV* mapping at -0.1 V is given in Fig. 3c, together with the corresponding fast Fourier transform (FFT) image (Fig. 3d). When the bias is below -400 meV, only an unidirectional stripe pattern is visible at the *dI/dV* map, manifesting an interference with a reciprocal lattice vector $\vec{g_b}$. With the increment of the bias, QI patterns show the interference with reciprocal lattice vectors not only along the Γ-X direction, but also for the Γ-Y direction. The intensity of interference pattern with the reciprocal lattice vector $\vec{g_b}$ reaches the maximum at -200 meV, while with $\vec{g_a}$ has a maximum at around -100 meV, as illustrated in Fig. 3e. By comparing with the integrated DOS from our ARPES measurement (Fig. 3f), a coincidence was found, showing that the interference maxima for reciprocal lattice vectors always occur at the top/bottom of downwards γ, ε bands or upwards band α. Such coincidence manifests those downwards/upwards bands are spin-split, namely opposite spin direction on the opposite side of Γ in momentum space. Our observation of such "spin texture" agrees with Jiang *et al.*'s circular dichroism (CD) measurement[8].

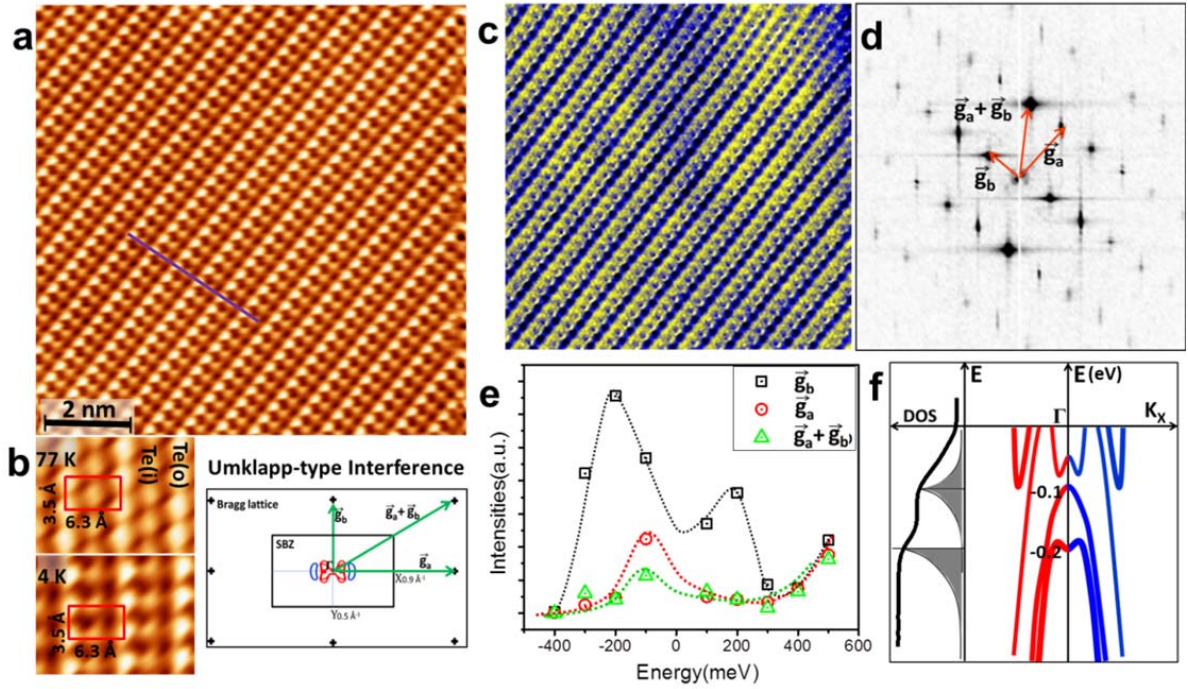

**Figure 3 | STM images and tunneling spectra at 4.2 K. (a)** STM topographic image, $V_b$ =500 mV, $I$=100 pA. **(b)** (Right) Schematic drawing showing umklapp-type interference processes in spin-conserving processes, involving a reciprocal lattice vector $\vec{g}$, are permitted, while ordinary interferences are forbidden by spin. (Left) A comparison between STM images taken at 77 K and 4.2 K. **(c)** $dI/dV$ conductance image with $V_b$ =-100 mV, $I$=100 pA. The size of image is 10 nm×10 nm. **(d)** FFT image of the $dI/dV$ image in panel **(c)**, showing umklapp-type interference, involving reciprocal lattice vectors $\vec{g_b}$, $\vec{g_a}$ and $\vec{g_b}+\vec{g_a}$. **(e)** The energy dependence of the intensities for interference spots, *i.e.* vectors $\vec{g_b}$, $\vec{g_a}$ and $\vec{g_b}+\vec{g_a}$, respectively. **(f)** A schematic illustration for bands near $E_F$ along Γ-X direction with the integrated DOS (thick black line) from ARPES data.

Now we can explain the distinct difference between STM topographies at 77 K and 4.2 K. Due to the presence of the surface enhanced Rashba SOC, the origin of the degenerate band is shifted by $k_R$, but in opposite directions for up- and down-spins with in overall spin-orbit lowering of $\Delta_R$ energy. This Rashba type spin orbit splitting is k-dependent, but is about 50 meV at some k, and lower at others. When the measuring temperature $T_m$ is raised *i.e.* at 77 K, the peculiar interference pattern vanishes and an STM image representing the surface geometry is observed meaning that scattering processes have sufficient energy mix bands at different energy, removing the spin protection from scattering; conversely at low T, the STM image will be dominated by such interference pattern, as what we observed at 4.2 K.

To further study the electronic nature of the surface, a spectroscopic survey was taken, consisting of 50 differential tunneling conductance spectra (*dI/dV* versus $V_b$). Figure 4a shows the *dI/dV* spectra acquired within five UCs (marked by a blue line in Fig. 3a). All curves exhibit similar line shapes without evidence for an energy gap. The spatial mapping of the differential tunneling conductance is shown in Fig. 4b. From this plot, oscillation of DOS of both filled states (below $E_F$) and empty states (above $E_F$) with lateral displacement, is clearly observed. Interestingly, the maximum position of filled states does not meet with the maximum of empty states. The separation between the lateral positions of the maxima for empty and filled states is about 2 Å, manifesting a separation of electron/hole-like channels.

It has been proposed that particle-hole asymmetry arises from an imbalance between the tunneling rate for electron injection and extraction[31,32], and this mechanism has been widely applied in the high-Tc cuprate superconductors studies[33,34]. According to the theory of spectral-weight sum rules[32], the ratio of particle-hole asymmetry $R(\vec{r})$ can be given by $R(\vec{r}) = \frac{\int_0^{E_2} N(\vec{r},E)dE}{\int_{-E_1}^{0} N(\vec{r},E)dE}$. By integrating empty and filled states from -0.2 V to 0 and from 0 to 0.4 V, respectively, a normalized particle/hole asymmetry ratio is plotted, as shown in Fig. 4c. The electron/hole ratio is varying within a unit cell along $b$ axis, which is perpendicular to tungsten chain. In half of the unit cell, it is more electron-like with the ratio lower than one and vice versa in another half. The overall electron/hole ratio within a unit cell always equals to one, suggesting the perfect n-p charge compensation in $WTe_2$[1,3]. As far as we know, such a microscopic picture of separated electron/hole channels for a double-carrier system has not been observed previously for semimetallic materials.

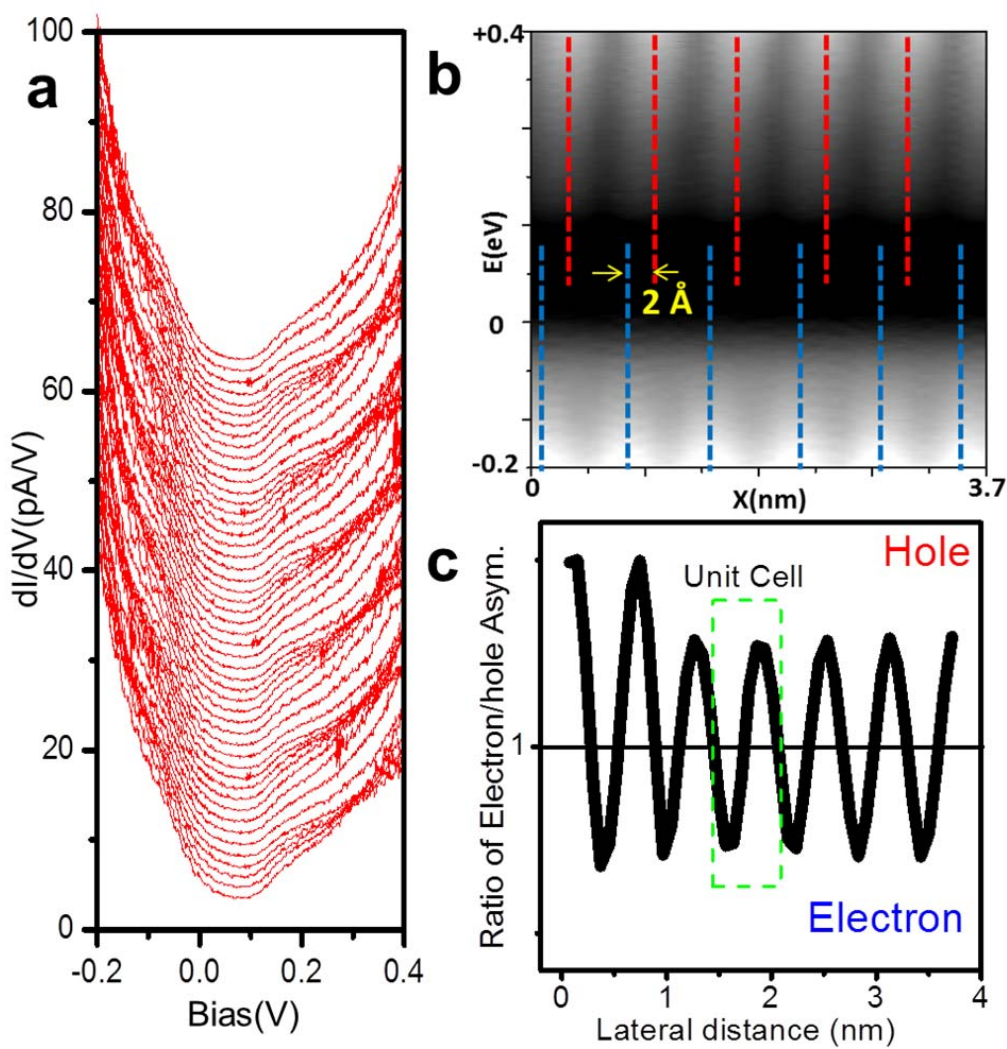

**Figure 4 | Low Temperature (4.2 K) differential conductance (*dI/dV*) spectra showing electron/hole asymmetry within a unit cell.** **(a)** A series of *dI/dV* spectra acquired along the blue line in Fig. 3a. Here, all spectra were taken with a bias of 50 mV, a tunneling current of 0.1 nA and The bias-modulation was set to 5 mV$_{rms}$. **(b)** A gray-scale plot of these *dI/dV* curves with the relative displacement. **(c)** The calculated ratio of electron/hole asymmetry based on the spectra survey in panel **(a)**.

Finally, we checked the electronic properties of WTe$_2$ surface under applied magnetic field. As shown in Fig. 5a, no distinct change in the spectrum is observed upon applying magnetic fields from 0 to 8 T. One possible reason for not observing Landau levels (LLs) is that the spacing of the Landau levels $\hbar\omega_c$ is too small to be measured. In contrast to the case of Cd$_3$As$_2$, in which the high Fermi velocity (~1×10$^6$ m/s) gives a spacing of approximately 20-30 meV between LLs, the largest spacing of Landau Levels of WTe$_2$ at 8 Tesla is estimated about 1-2 meV. Another possibility for not observing LLs is an extremely small cross-sectional area $S_n$ in reciprocal space. According to the reported ARPES work[7], the size of the electron-like and the hole-like pocket at the Fermi level are almost exactly the same in area, about 0.018 Å$^{-2}$. This similarity as well as the overall small size of the Fermi surfaces causes tiny cross-sectional area perpendicular to the magnetic field, therefore may play an important role in the non-observation of quantized Landau levels at moderate field. A spectroscopic survey by applying 8 Tesla perpendicular field, consisting of 50 dI/dV spectra crossing five UCs is acquired, as shown in Fig. 5b. The similarity between Fig. 4b and Fig. 5b suggests the existence of a separated electron/hole-like channels at 8 T. In the other words, one may notice that the spatial oscillations of electron injection and extraction become more blurred at high magnetic field. This effect could be due to Fermi's golden rule, in which the scattering rate between one electron and one hole LL increases proportionally with the magnetic field.

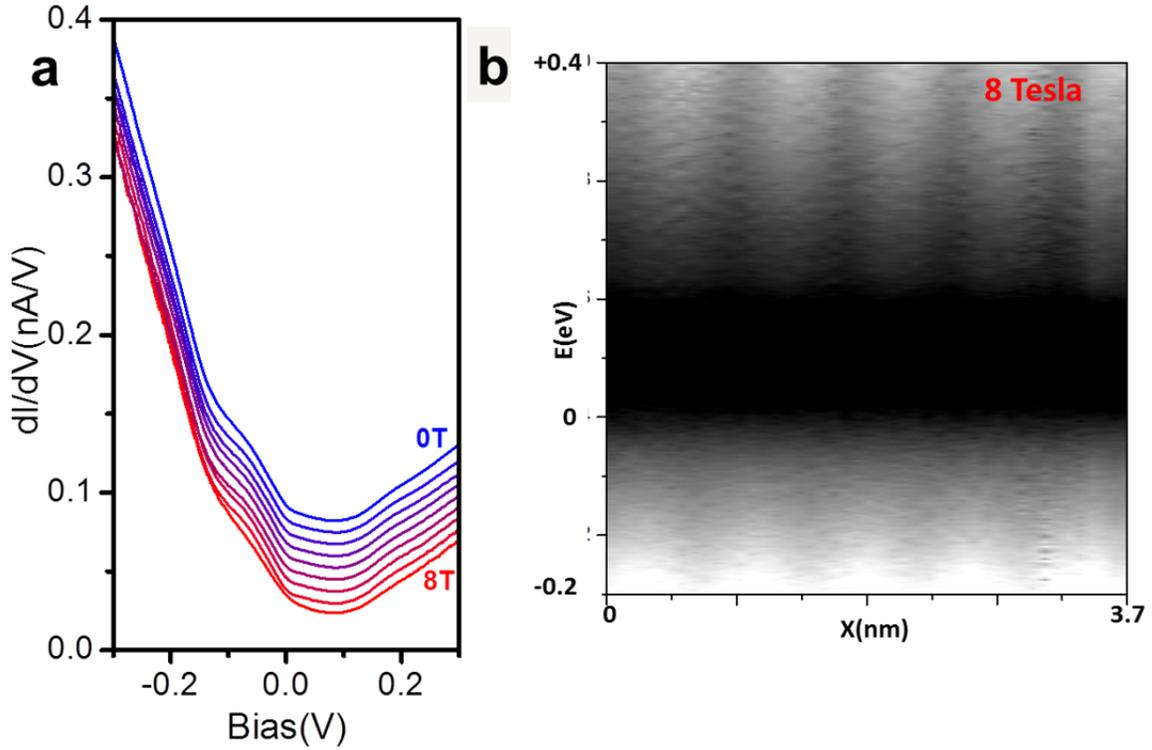

**Figure 5 | Tunneling spectra under a magnetic field along *c* axis.** (a) *dI/dV* spectra acquired at different perpendicular fields (from top to bottom: 0, 1, 2, 3, 4, 5, 6, 7, 8 Tesla, respectively). Here, all spectra were taken at 4.2 K with $V_b$=50 mV and $I$= 0.1 nA. Bias-modulation amplitude was set to 5 mV$_{rms}$. (b) A color plot of the *dI/dV* curves with the relative distance, from a line spectra survey taken under 8 Tesla field.

## Discussion

In summary, utilizing STM/S, ARPES and transport measurements, corroborated by DFT calculations, we revealed that strong Rashba type spin-orbit effects due to the non-centrosymmetric structure splits the electron and hole bands crossing the Fermi level and lifts the spin degeneracy of these bands. Such effect almost doubles the number of Fermi pockets, producing five bands that cross $E_F$, instead of two bands normally. We also note that the Fermi surfaces and their topology found both experimentally and from our DFT calculations differ from those initially reported[1]. Furthermore, low temperature STM/S measurements show the umklapp-type interference pattern on $WTe_2$ surface at 4.2 K, proven to be a spectroscopic feature instead of a pure geometric one, which is dominated by spin-conserving processes. Spectroscopic surveys reveal that the ratio of electron/hole asymmetry changes alternately with lateral locations along *b* axis, while the overall electron/hole ratio within a unit cell stays into a perfect balance. In $WTe_2$ the Fermi surface is arranged along the a-axis which is the current direction for the XMR and the direction in which the Rashba-type splittings are important. The inversion symmetry is broken by a lack of reflection in the c-direction. This leads to an important spin texture in the *a-b* plane, with polarized tangential to the Fermi surface. Thus the magnetic field along *c* breaks this texture. We infer that it is this that enables scattering between parts of the Fermi surface that have different spin without field and underlies the XMR effect. Our findings provide a microscopic atomic level understanding of the electronic and spin structure of the exotic semimetal material $WTe_2$ and show the importance of the interplay of spin-orbit interactions with the non-centrosymmetric crystal structure.

## Methods

**Synthesis.** Single crystals of WTe$_2$ are grown via self-flux technique. W and Te shots in an atomic ratio of 1:49 were placed in a 5 ml Al$_2$O$_3$ crucible to grow the WTe$_2$ single crystals. A catch crucible containing quartz wool was mounted on top of growth crucible and both were sealed in a silica ampoule under approximately 1/3 atmosphere of high pure argon gas. The sealed ampoule was heated up initially to 1100 °C and kept for 6 hours, then cooled down to 500 °C over 96 hours. The Te flux was separated from single crystals by using a centrifuge once the temperature reaches 500 °C.

**STM/S.** For the STM/S measurements, a Unisoku low-temperature scanning tunneling microscopy was used for the imaging and spectroscopic measurements. The sample was cleaved at room temperature in ultra-high vacuum to expose a shining surface and then loaded into the STM head for investigation at 77 K and 4.2 K, respectively. We obtained topographic images in constant-current mode, and the tunneling spectra *dI/dV* using lock-in technique to measure differential conductance. A chemically etched W tip was ued. The bias voltage was applied on the sample side during the STM observations. The WSxM software has been used to process and analyze STM data.

**DFT calculations.** Density functional theory calculation were done using the generalized gradient approximation of Perdew, Burke and Ernzerhof (PBE)[35]. We used the general potential linearized augmented planewave method[36] as implemented in the WIEN2k code[37] for calculations of the electronic structure and for structure relaxation. We used well converged basis sets consisting of standard LAPW functions up to a cutoff RK$_{max}$=9, where K$_{max}$ is the planewave sector cutoff and R is the minimum LAPW sphere radius, in this case 2.45 bohr; the sphere radii were 2.45 Bohr for W and 2.55 Bohr for Te. Additionally, local orbitals were added for the W 4f, 5s and 5p and Te 4d semicore states. Relativity was included at the Dirac level for the core states and in a scalar relativistic approximation for the valence states in the structure relaxation. Once the relaxation was done we calculated the electronic band structure, Fermi surfaces, plasma frequencies and properties including spin orbit.

## Acknowledgements

STM work was conducted at Jiangsu Key Laboratory for Carbon-based Functional Materials & Devices, Soochow University. Part of research was supported by the U.S. Department of Energy, Basic Energy Sciences, Materials Sciences and Engineering Division. QL and LFC acknowledge the financial supported by the Major State Basic Research Development Program of China (2014CB932600), National Natural Science Foundation of China (Project Code 91227201 and 21403149) and Natural Science Foundation of Jiangsu Province (BK20140305). DGM acknowledges support from the Gordon and Betty Moore Foundation's EPiQS Initiative through Grant GBMF4416. JQY acknowledges support from the US Department of Energy, Office of Science, Basic Energy Sciences, Materials Sciences and Engineering Division. Work at the University of Missouri (DJS) is supported by the Department of Energy, Office of Science, Basic Energy Sciences through the Computational Synthesis of Materials Software Project with Validation on Layered Low Dimensional Functional Materials and Ultra-Fast X-Ray Laser Experiments.

## Author contributions